# Solution to two-dimensional Incompressible Navier-Stokes Equations with SIMPLE, SIMPLER and Vorticity-Stream Function Approaches. Driven-Lid Cavity Problem: Solution and Visualization.

by

**Maciej Matyka**

Computational Physics Section of Theoretical Physics
University of Wrocław in Poland
Department of Physics and Astronomy

Exchange Student at University of Linköping in Sweden

maq@panoramix.ift.uni.wroc.pl
http://panoramix.ift.uni.wroc.pl/∼maq

30 czerwca 2004 roku

## Streszczenie

In that report solution to incompressible Navier - Stokes equations in non - dimensional form will be presented. Standard fundamental methods: SIMPLE, SIMPLER (SIMPLE Revised) and Vorticity-Stream function approach are compared and results of them are analyzed for standard CFD test case - Drived Cavity flow. Different aspect ratios of cavity and different Reynolds numbers are studied.

## 1 Introduction

The main problem is to solve two-dimensional Navier-Stokes equations. I will consider two different mathematical formulations of that problem:

- $u, v, p$ primitive variables formulation
- $\zeta, \psi$ vorticity-stream function approach

I will provide full solution with both of these methods. First we will consider three standard, primitive component formulations, where fundamental Navier-Stokes equation will be solved on rectangular, staggered grid. Then, solution on non-staggered grid with vorticity-stream function form of NS equations will be shown.

## 2 Math background

We will consider two-dimensional Navier-Stokes equations in non-dimensional form[1]:

[1] We consider flow without external forces i.e. without gravity.

$$\frac{\partial \overrightarrow{u}}{\partial t} = -(\overrightarrow{u}\nabla)\overrightarrow{u} - \nabla\varphi + \frac{1}{Re}\nabla^2\overrightarrow{u} \tag{1}$$

$$D = \nabla\overrightarrow{u} = 0 \tag{2}$$

Where equation (2) is a continuity equation which has to be true for the final result.

# 3 Primitive variables formulation

First we will examine SIMPLE algorithm which is based on primitive variables formulation of NS equations. When we say "primitive variables" we mean $u, v, p$ where $\mathbf{u} = (u, v)$ is a velocity vector, and $p$ is pressure. We can rewrite equation (1) in differential form for both velocity components:

$$\frac{\partial u}{\partial t} = -\frac{\partial u^2}{\partial x} - \frac{\partial uv}{\partial y} - \frac{\partial p}{\partial x} + \frac{1}{Re}(\frac{\partial^2 u}{\partial x^2} + \frac{\partial^2 v}{\partial y^2}) \tag{3}$$

$$\frac{\partial v}{\partial t} = -\frac{\partial v^2}{\partial y} - \frac{\partial uv}{\partial x} - \frac{\partial p}{\partial y} + \frac{1}{Re}(\frac{\partial^2 u}{\partial x^2} + \frac{\partial^2 v}{\partial y^2}) \tag{4}$$

We rewrite continuity equation in the following form:

$$\frac{\partial u}{\partial x} + \frac{\partial v}{\partial y} = 0 \tag{5}$$

These equations are to be solved with SIMPLE method.

## 3.1 SIMPLE algorithm

SIMPLE algorithm is one of the fundamental algorithm to solve incompressible NS equations. SIMPLE means: Semi Implicit Method for Pressure Linked Equations.

Algorithm used in my calculations is presented in the figure (1). First we have to guess initial values of the pressure field[2] $(P^*)^n$ and set initial value of velocity field - $(U^*)^n$, $(V^*)^n$. Then equation (3) and (4) is solved to obtain values of $(U^*)^{n+1}$, $(V^*)^{n+1}$. Next we have to solve pressure-correction equation:

$$\nabla^2 p' = \frac{1}{\Delta t}(\nabla \cdot V) \tag{6}$$

Next - a simple relation to obtain corrected values of pressure and velocity fields is applied (see appendix A for details about velocity correction calculaion). At the end of time step we check if solution coverged.

[2]Subscripts denote computational step, where "n+1" means current step.

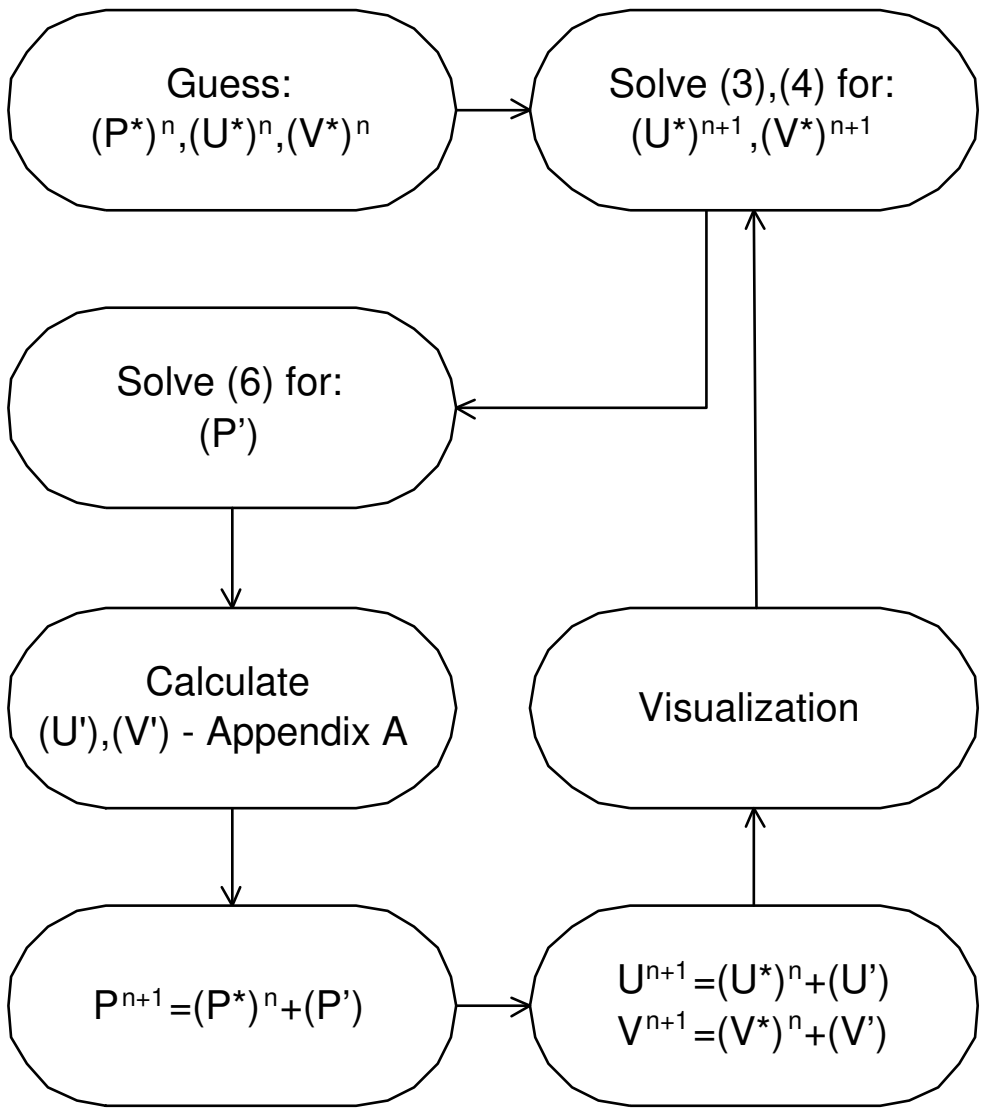

Rysunek 1: SIMPLE Flow Diagram

## 3.2 Numerical Methods in SIMPLE

### 3.2.1 Staggered Grid

For discretization of differential equations I am using staggered grid. In the figure (2) staggered grid for rectangular area is sketched. Primitive variables are placed in different places. In points $i, j$ on a grid pressure $P$ values, in points $i + 0.5, j$ $u$ x-velocity components and in points $i, j + 0.5$ $v$ y-velocity components are placed. That simple model of staggered grid gives us possibility to use simple discretization with second order accuracy which will be discussed later.

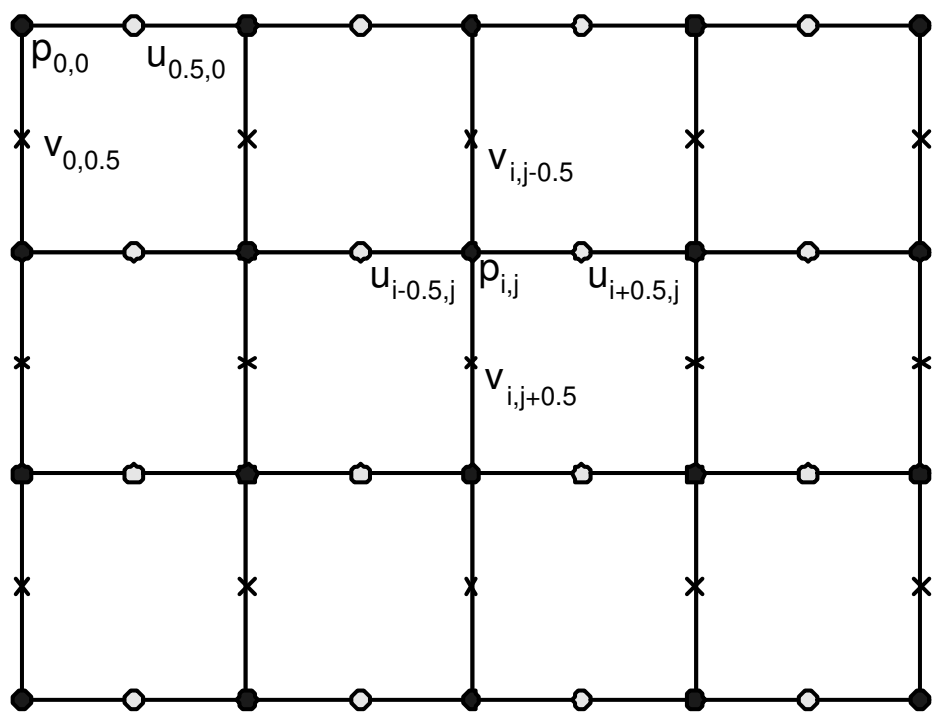

Rysunek 2: Staggered grid: filled circles $P$, outline circles $U$ x-velocity, cross $V$ y-velocity component.

| Differential | Discretization | Type |
|---|---|---|
| $\frac{\partial u}{\partial t}$ | $\frac{u^{n+1}-u^{n}}{\Delta t}$ | forward, $O(h)$ |
| $\frac{\partial^2 u}{\partial x^2}$ | $\frac{u_{i+1,j}-2*u_{i,j}+u_{i-1,j}}{(\Delta x)^2}$ | central, $O(h^2)$ |
| $\frac{\partial u^2}{\partial x}$ | $\frac{u^2_{i+1,j}-u^2_{i-1,j}}{(2\cdot\Delta x)}$ | central, $O(h^2)$ |
| $\frac{\partial p}{\partial x}$ | $\frac{p_{i+1,j}-p_{i,j}}{(\Delta x)}$ | forward, $O(h)$ |
| $\frac{\partial p}{\partial y}$ | $\frac{p_{i,j+1}-p_{i,j}}{(\Delta y)}$ | forward, $O(h)$ |

Tabela 1: Discretizations used in SIMPLE algorithm

#### 3.2.2 Discretization Schemes

Let us now examine some numerical methods used in presented solution. For algorithm presented in the figure (1) we have only three equations which have to be discretized on a grid. First we have momentum equations (3) and (4).

Discrete schemes used in discretization of momentum equations are presented in a table (1). Using presented discrete form of derivatives I obtain numerical scheme for momentum equations exactly in the form presented in [1]. Equations (3) , (4) discretized on staggered grid can be written[3] as follows[4]:

$$u^{n+1}_{i+0.5,j} = u^{n}_{i+0.5,j} + \Delta t \cdot (A - (\Delta x)^{-1}(p_{i+1,j} - p_{i,j})) \quad (7)$$

$$v^{n+1}_{i,j+0.5} = v^{n}_{i,j+0.5} + \Delta t \cdot (B - (\Delta y)^{-1}(p_{i,j+1} - p_{i,j})) \quad (8)$$

where $A$ and $B$ are defined as:

$$A = -a_1 + (Re)^{-1} \cdot (a_3 + a_4) \quad (9)$$

$$B = -b_1 + (Re)^{-1} \cdot (b_3 + b_4) \quad (10)$$

and respectively we define:

$$a_1 = -\frac{(u^2)^n_{i+1.5,j} - (u^2)^n_{i-0.5,j}}{2 \cdot \Delta x} - \frac{(u\dot{v})^n_{i+0.5,j+1} - (u\dot{\dot{v}})^n_{i+0.5,j-1}}{2 \cdot \Delta y} \quad (11)$$

$$b_1 = -\frac{(v^2)^n_{i,j+1.5} - (v^2)^n_{i,j-0.5}}{2 \cdot \Delta y} - \frac{(v\dot{u})^n_{i+1,j+0.5} - (v\dot{\dot{u}})^n_{i-1,j+0.5}}{2 \cdot \Delta x} \quad (12)$$

$$(a_3) = \frac{u^n_{i+1.5,j} - 2 \cdot u^n_{i+0.5,j} + u^n_{i-0.5,j}}{(\Delta x)^2} \quad (13)$$

$$(a_4) = \frac{u^n_{i+0.5,j+1} - 2 \cdot u^n_{i+0.5,j} + u^n_{i+0.5,j-1}}{(\Delta y)^2} \quad (14)$$

$$(b_3) = \frac{v^n_{i,j+1.5} - 2 \cdot v^n_{i,j+0.5} + v^n_{i,j-0.5}}{(\Delta y)^2} \quad (15)$$

$$(b_4) = \frac{v^n_{i+1,j+0.5} - 2 \cdot v^n_{i,j+0.5} + v^n_{i-1,j+0.5}}{(\Delta x)^2} \quad (16)$$

Now we have defined almost everything. Dotted velocities should be also defined. I use simple expressions for it:

$$\dot{u} = 0.5 \cdot (u_{i-0.5,j} + u_{i-0.5,j+1}) \quad (17)$$

$$\dot{\dot{u}} = 0.5 \cdot (u_{i+0.5,j} + u_{i+0.5,j+1}) \quad (18)$$

$$\dot{v} = 0.5 \cdot (v_{i,j+0.5} + v_{i+1,j+0.5}) \quad (19)$$

$$\dot{\dot{v}} = 0.5 \cdot (v_{i,j-0.5} + v_{i+1,j-0.5}) \quad (20)$$

#### 3.2.3 Poisson Equation

For equation I use simple iterative procedure. In the figure (3) points used for calculation of pressure at each $(i, j)$ grid points are marked.

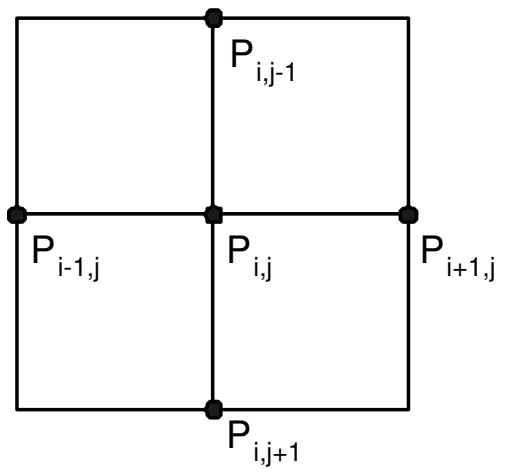

Rysunek 3: Points on a grid used in iterative procedure for Poisson equation solving.

I use simple 4 points scheme for Laplace operator. Directly from [1] expression for one iterative step of poisson equation solver can be written as follows:

$$p'_{i,j} = -a^{-1}(b \cdot (p'_{i+1,j} + p'_{i-1,j}) + c \cdot (p'_{i,j+1} + p'_{i,j-1}) + d) \quad (21)$$

[3] Please note than cited [1] reference contains some print mistakes there.

[4] Generally I show there only an idea how to write discretized equations, they should be rewritten with "*" and "'" chars for concrete steps of the algorithm

where

$$a = 2\Delta t \left[ \frac{1}{\Delta x^2} + \frac{1}{\Delta y^2} \right] \quad (22)$$

$$b = -\frac{\Delta t}{\Delta x^2} \quad (23)$$

$$c = -\frac{\Delta t}{\Delta y^2} \quad (24)$$

$$d = \frac{1}{\Delta x} \left[ u_{i+0.5,j} - u_{i-0.5,j} \right] + \frac{1}{\Delta y} \left[ v_{i,j+0.5} - v_{i,j-0.5} \right] \quad (25)$$

That iterative procedure is rather simple - we use equation (21) for all interior points on a grid. After that one step of iterative procedure is done. Then we check if solution coverges. We can do it simply to check maximum change of pressure on a grid. If it is bigger than $\epsilon$ we continue iterative process. Solution should finish when pressure field is exactly coverged $\epsilon = 0$, but in practice I use different value of $\epsilon$ for different physical properties of simulated models - it will be discussed later.

### 3.3 SIMPLE Revised algorithm

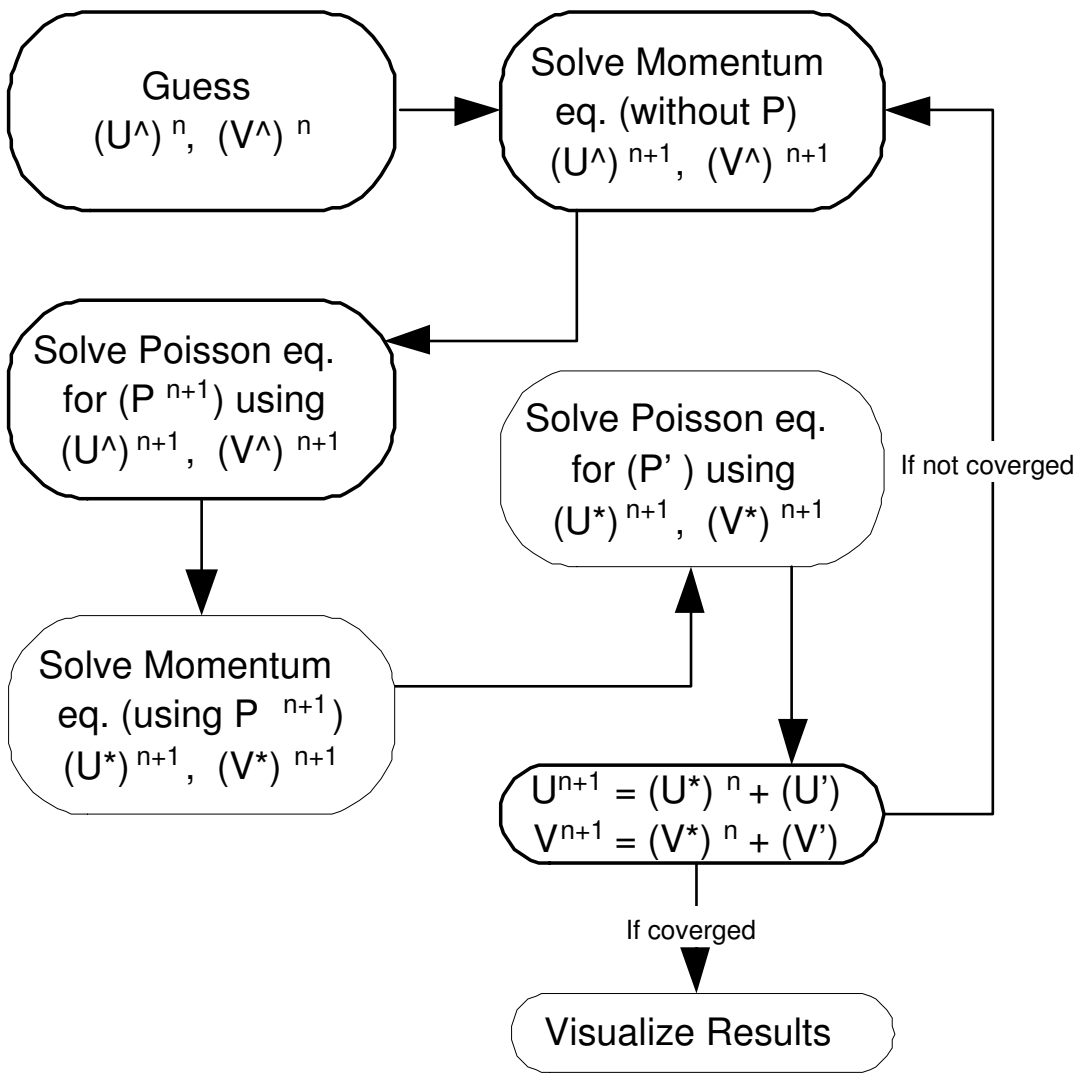

Rysunek 4: Flow chart for SIMPLER algorithm.

In the figure (4) I present SIMPLE Revised algorithm. It is easy to extend existing SIMPLE solution to be SIMPLER one.

Treating the boundary conditions and numerical methods used in SIMPLER solution is almost the same as in SIMPLE, so I will not repeat myself.

## 4 Vorticity-Stream Function approach

Vorticity-Stream Function approach to two-dimensional problem of solving Navier-Stokes equations is rather easy. A different form of equations can be scary at the beginning but, mathematically, we have only two variables which have to be obtained during computations: stream vorticity vector $\zeta$ and stream function $\Psi$.

First let us provide some definition which will simplify NS equation. The main goal of that is to remove explicitly Pressure from N-S equations. We can do it with the procedure as follows.

First let us define vorticity for 2D case:

$$\zeta = |\zeta| = |\nabla \times V| = \frac{\partial v}{\partial x} - \frac{\partial u}{\partial y} \quad (26)$$

And stream function definition is:

$$\frac{\partial \Psi}{\partial y} = u \quad (27)$$

$$\frac{\partial \Psi}{\partial x} = -v \quad (28)$$

We can combine these definitions with equations (3) and (4). It will eliminate pressure from these momentum equations. That combination will give us non-pressure vorticity transport equation which in non-steady form can be written as follows:

$$\frac{\partial \zeta}{\partial t} + u\frac{\partial \zeta}{\partial x} + v\frac{\partial \zeta}{\partial y} = \frac{1}{Re}\left(\frac{\partial^2 \zeta}{\partial x^2} + \frac{\partial^2 \zeta}{\partial y^2}\right) \quad (29)$$

Having combined equations (26), (27) and (28) we obtain poisson equation for the $\Psi$ variable:

$$\nabla^2 \Psi = \frac{\partial^2 \Psi}{\partial x^2} + \frac{\partial^2 \Psi}{\partial y^2} = -\zeta \quad (30)$$

Now we have all definitions and equations which are needed for vorticity-stream solution. We will solve vorticity transport equation, then new values of $\zeta$ will be used to solve equation (30).

### 4.1 Non-Staggered Grid

Instead of using staggered grid in Vorticity-Stream approach, we will place both $\zeta$ and $\Psi$ variables in the same place as it is shown in the figure (5)

Discretization is straightforward and easier to implement in a non-staggered grid than in a staggered grid for the SIMPLE algorithm.

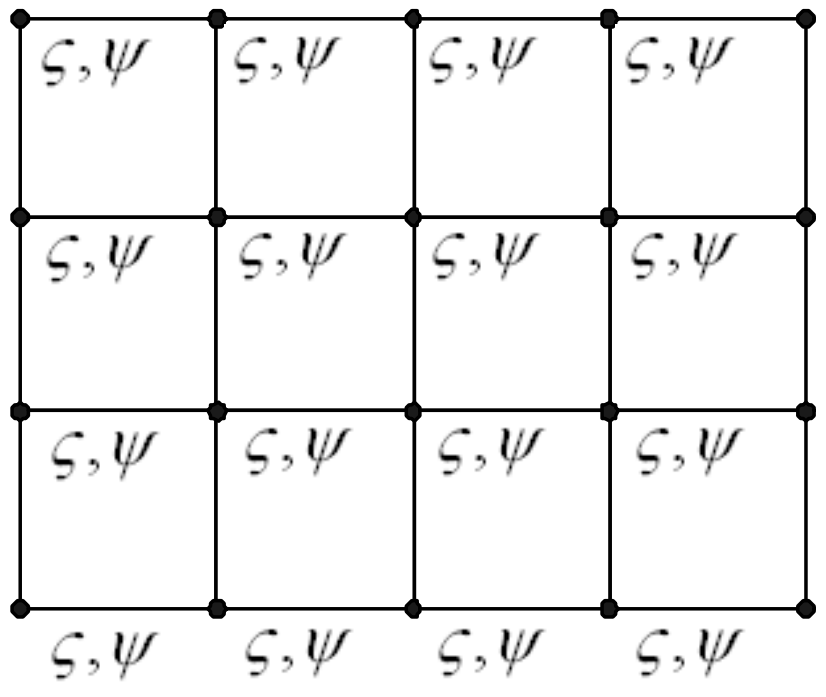

Rysunek 5: $\zeta$ and $\Psi$ variables in non staggered grid.

| Differential | Discretization | Type |
|---|---|---|
| $\frac{\partial \zeta}{\partial t}$ | $\frac{\zeta^{n+1}-\zeta^{n}}{\Delta t}$ | forward, $O(h)$ |
| $\frac{\partial^2 \zeta}{\partial x^2}$ | $\frac{\zeta_{i+1,j}-2*\zeta_{i,j}+\zeta_{i-1,j}}{(\Delta x)^2}$ | central, $O(h^2)$ |
| $\frac{\partial \zeta}{\partial x}$ | $\frac{\zeta_{i+1,j}-\zeta_{i-1,j}}{(2\cdot\Delta x)}$ | central, $O(h^2)$ |
| $\frac{\partial^2 \Psi}{\partial x^2}$ | $\frac{\Psi_{i+1,j}-2*\Psi_{i,j}+\Psi_{i-1,j}}{(\Delta y)^2}$ | central, $O(h^2)$ |

Tabela 2: Discretizations used in Vorticity-Stream algorithm

### 4.2 Discretization

We will use several schemes to discretize differential equation (26). For Poisson equation we will use the same technique which was presented in the SIMPLE algorithm description, so we will not repeat formulas[5].

### 4.3 Vorticity-Stream function algorithm

Algorithm of solution for VS function solution is simplier than for SIMPLE method. It is sketched in the figure (6).

First we have to set initial values for $\zeta$ and $\Psi$. I arbitrary set these values to 0. Then Vorticity Transport Equation is solved and new $\zeta^{n+1}$ values are obtained. After that simple iterative procedure is applied to solve the poisson equation. Finally, new values of velocities are easily found from (27) and (28) equation.

[5]Formulas for poisson equation will be a little bit different but it is rather easy to obtain it by simple discretization of equation (30).

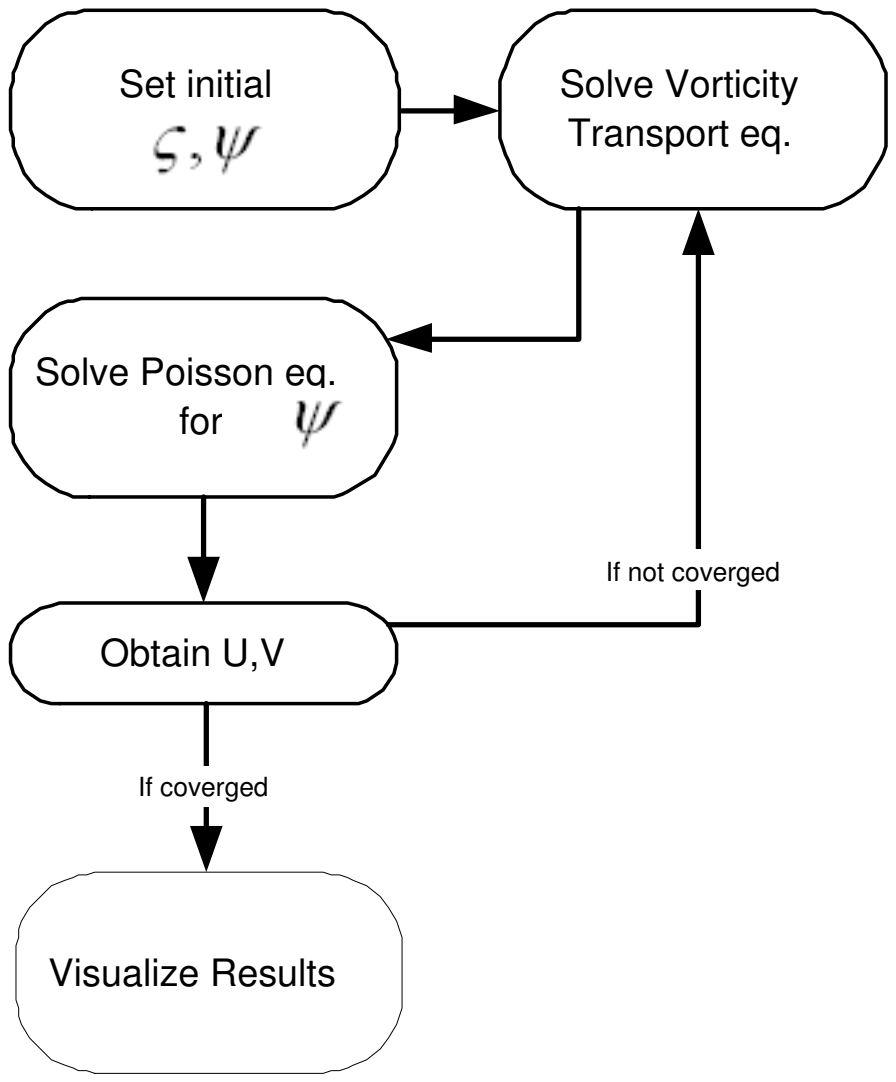

Rysunek 6: Algorithm of Vorticity-Stream solution.

## 5 Two-dimensional Driven-Lid Cavity

Let us now provide some examples of practical calculation for implemented methods[6]. I will show results of Driven-Lid Cavity flow - a standard CFD test case to check the solution.

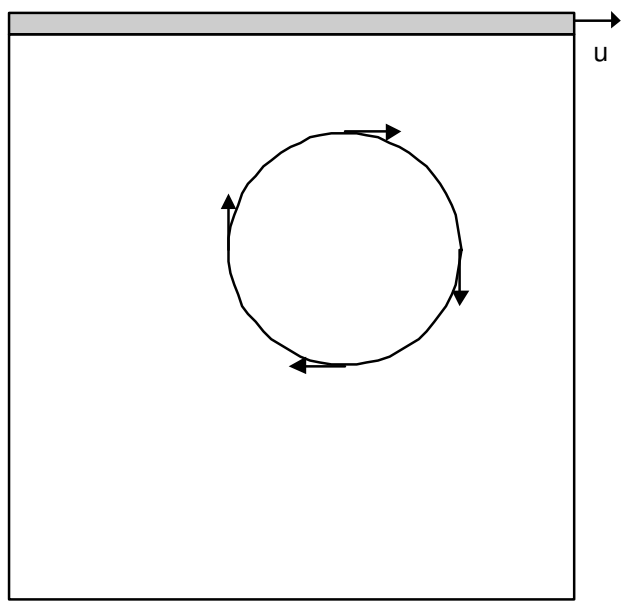

Rysunek 7: Driven Cavity (lid moving with $u$ constants velocity.

Driven Cavity problem is sketched in the figure (7). Upper lid is moving with $u$ velocity. Main goal is to solve NS equations inside the cavity to obtain velocity field (steady state). First of all we have to decide about boundary conditions for both: SIMPLE and VS approaches which will be quite different.

[6]In that section also boundary conditions will be provided, because they are specified especially for the given problem.

### 5.1 Boundary Conditions - SIMPLE and SIMPLER

For SIMPLE(R) method we will use BC as follows: First we have to clear pressure values for all boundaries. We use simple expression:

$$\frac{\partial p}{\partial \overline{n}} = 0 \tag{31}$$

where $\overline{n}$ is normal to the wall. It means that for all $i = 0 \ldots NX-1$ points of a grid we apply:

$$p_{i,0} = p_{i,1} \tag{32}$$

and

$$p_{i,ny-1} = p_{i,ny-2} \tag{33}$$

We apply that procedure for upper and lower wall respectively[7]. Then we have to take care of velocities. We would like to apply NOSLIP boundaries for Driven Cavity non-moving walls, so we have to zero values of velocities on every wall. First let us make trivial operation: for every $j = 0 \ldots NY-1$ set

$$v_{0,j} = 0 \tag{34}$$

and

$$v_{nx-1,j} = 0 \tag{35}$$

The same work should be done for $u$ velocities, for $i = 0 \ldots NX-1$ and for $j = NY-1$. Especially for driven cavity problem we also have to set $u$ velocity equal to 1.0 at $j = 0$ row, which is done in a straightforward way. One problem is to set boundary conditions at other walls, where no velocity grid points are present. We can do it with a simple linear interpolation of near velocities i.e. for $u$ velocity, for every $j = 0 \ldots NY-1$ we set:

$$u_{0,j} = -(2.0/3.0) \cdot u_{1,j} \tag{36}$$

and

$$u_{nx-2,j} = -(2.0/3.0) \cdot u_{nx-3,j} \tag{37}$$

The same condition is used for other walls and $v$ velocity components.

[7] For corners simple diagonal values are taken, i.e. $p_{0,0} = p_{1,1}$

### 5.2 Boundary Conditions - Vorticity Stream

In vorticity-stream formulation I use simple first order expressions for $\zeta$ derivatives at the wall. First, we have to set $\Psi = 0$ at all boundaries. Then for NOSLIP boundary walls we use expression (i.e. for $j = ny-1$ row):

$$\zeta_{i,0} = 2.0 \cdot \frac{\Psi_{i,0} - \Psi_{i,1}}{\Delta y^2} \tag{38}$$

## 6 Results

In that section some numerical results of calculations with three different techniques will be presented. Since results of calculations are the same I will try to show and compare differences between methods (accuracy, convergence speed). Please note that all comments are under figures.

### 6.1 Vorticity-Stream, Driven Cavity, $Re = 500$, Grid: 40x40

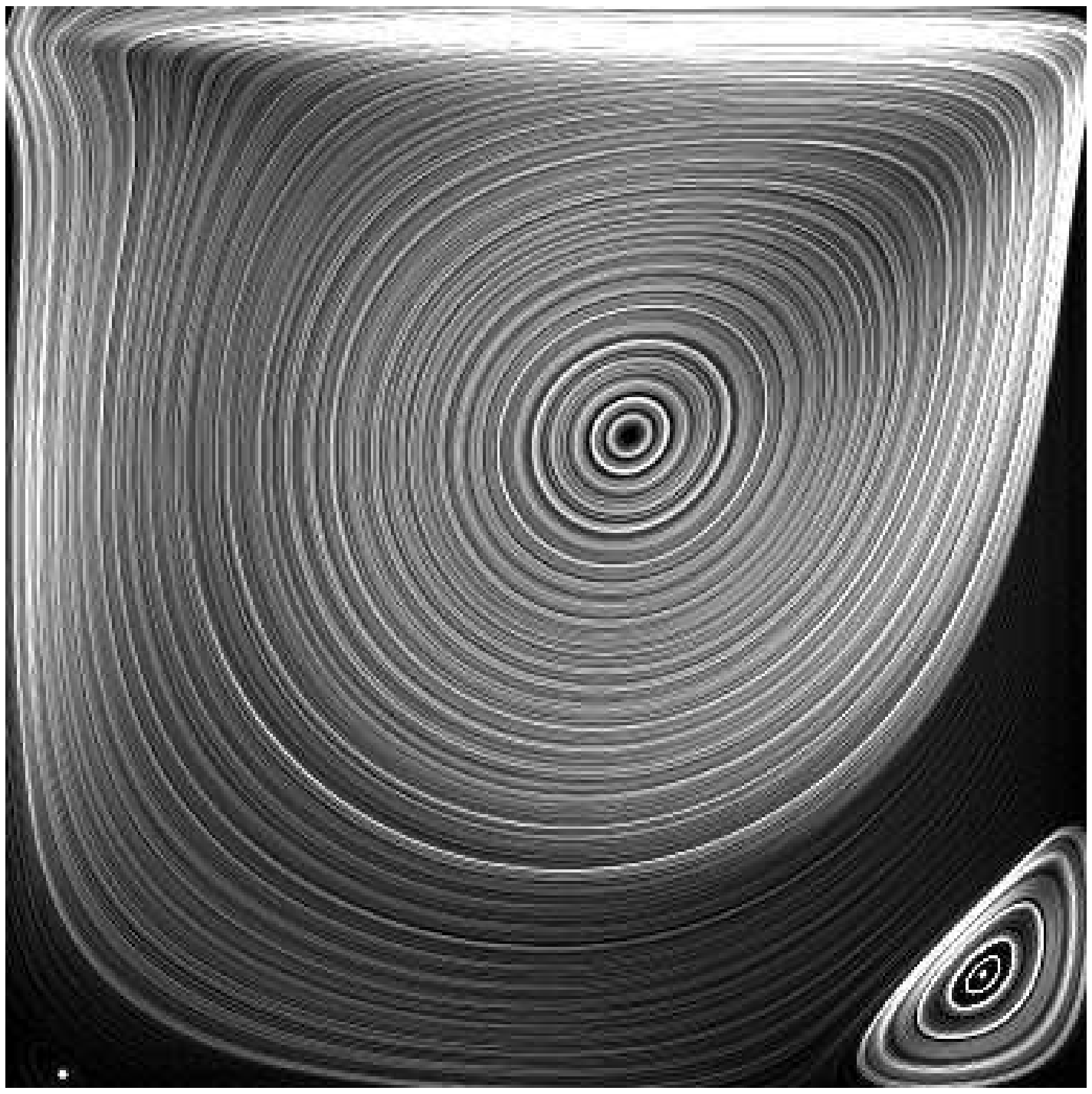

Rysunek 8: Streamlines plot for driven cavity with $Re = 500$ and $1:1$ aspect ratio, grid size $40x40$. Two vortexes are found in the corners of the Cavity, computed with the Vorticity-Stream approach. Solution visualized with Streamline plot technique.

### 6.2 Different Visualization Techniques, Driven Cavity, $Re = 500$, Grid: 40x40

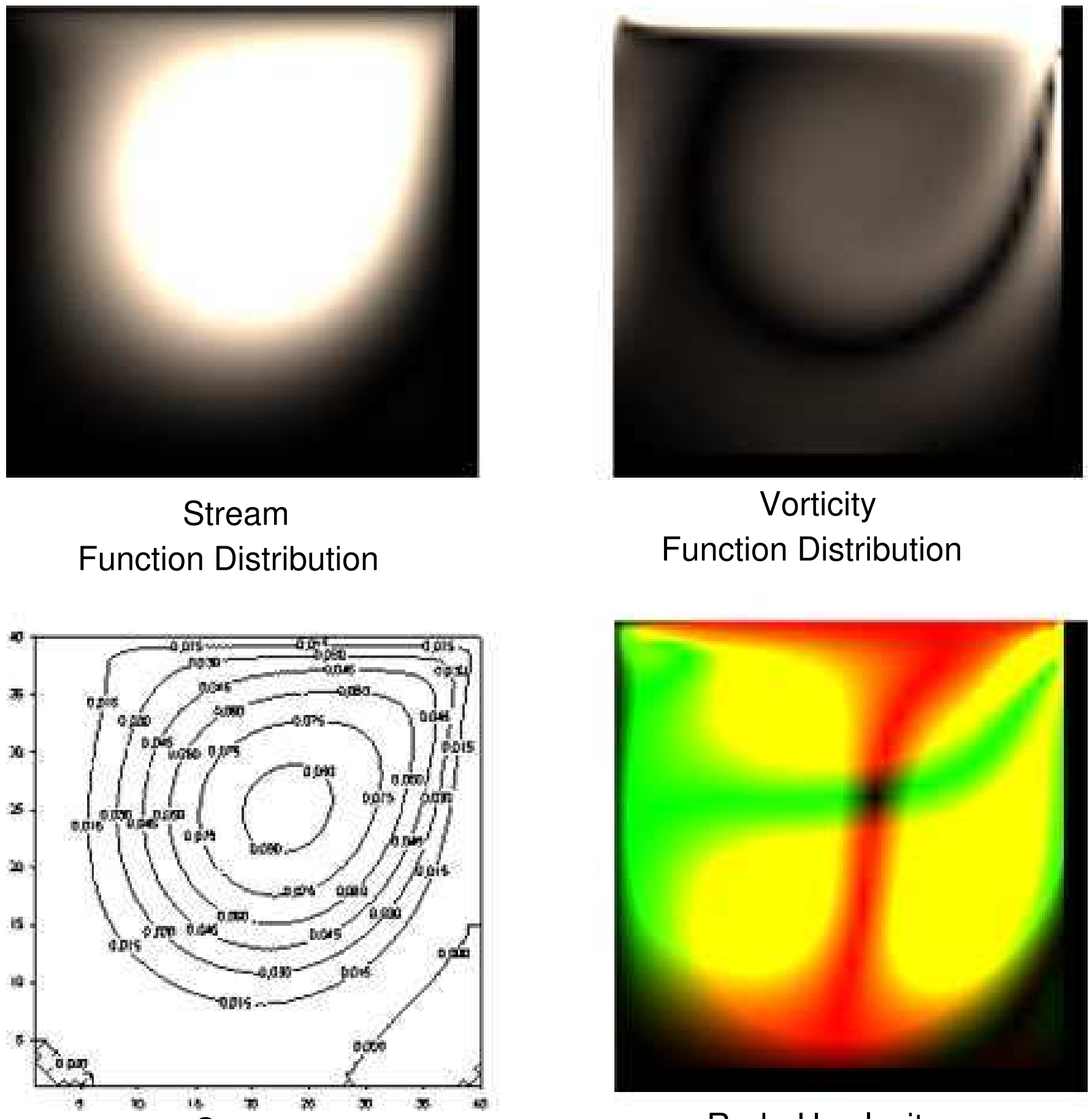

Stream
Function Distribution

Vorticity
Function Distribution

Stream
Function Contour Plot

Red - U velocity
Green - V velocity

Rysunek 9: There are presented different types of visualizations generated by my solver. Computations as above - $Re = 500$ and other parameters are the same. (That is only a part of possibility visualizations, more will be available on my web page soon).

### 6.3 SIMPLE, SIMPLER, Driven Cavity, $Re = 100$, Grid: 21x21

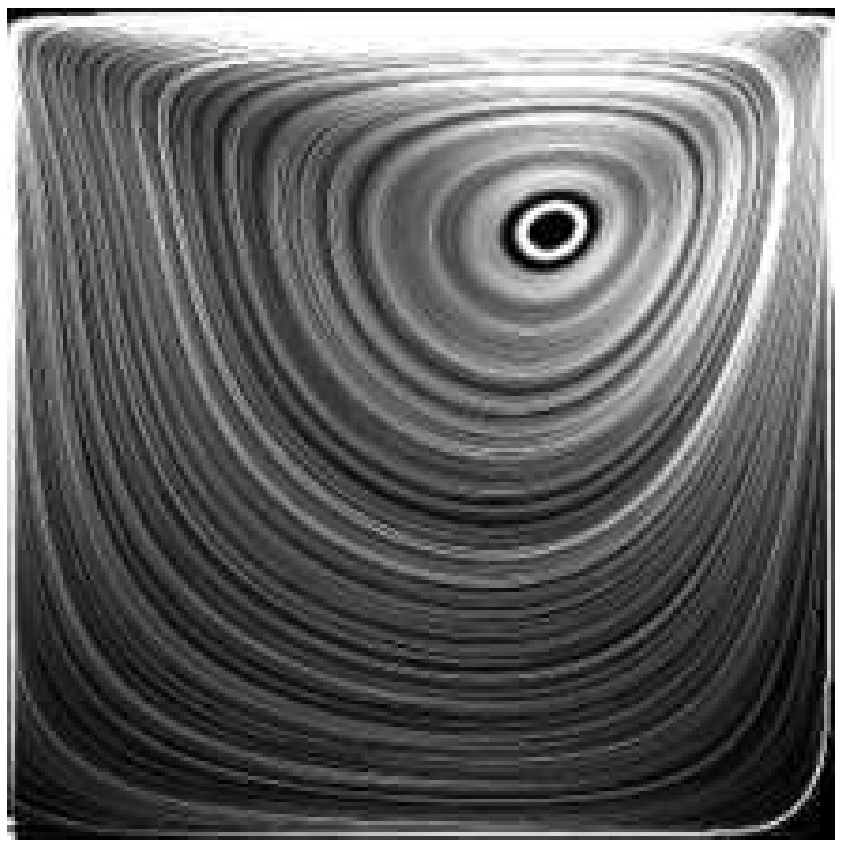

40x40, Aspect Ratio 1:1

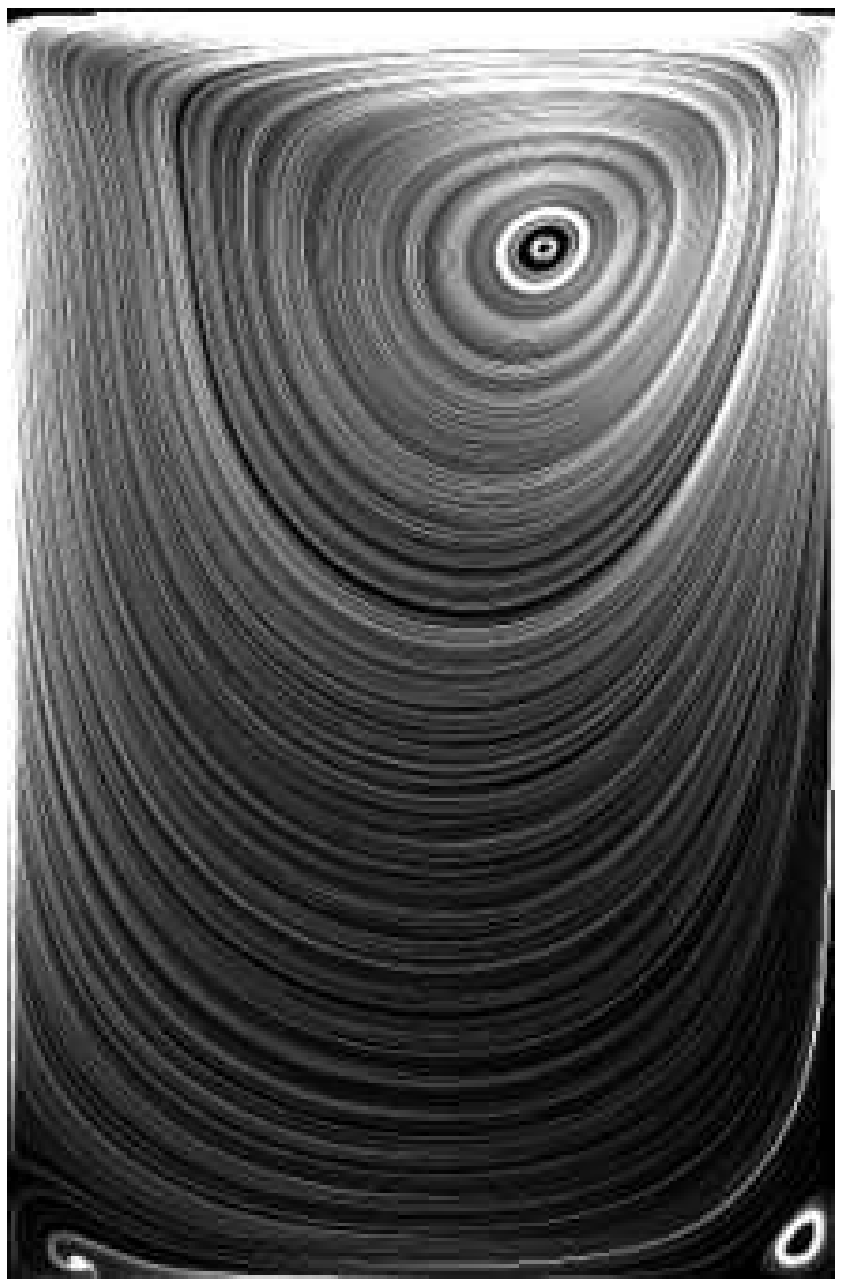

40x60, Aspect Ratio 2:3

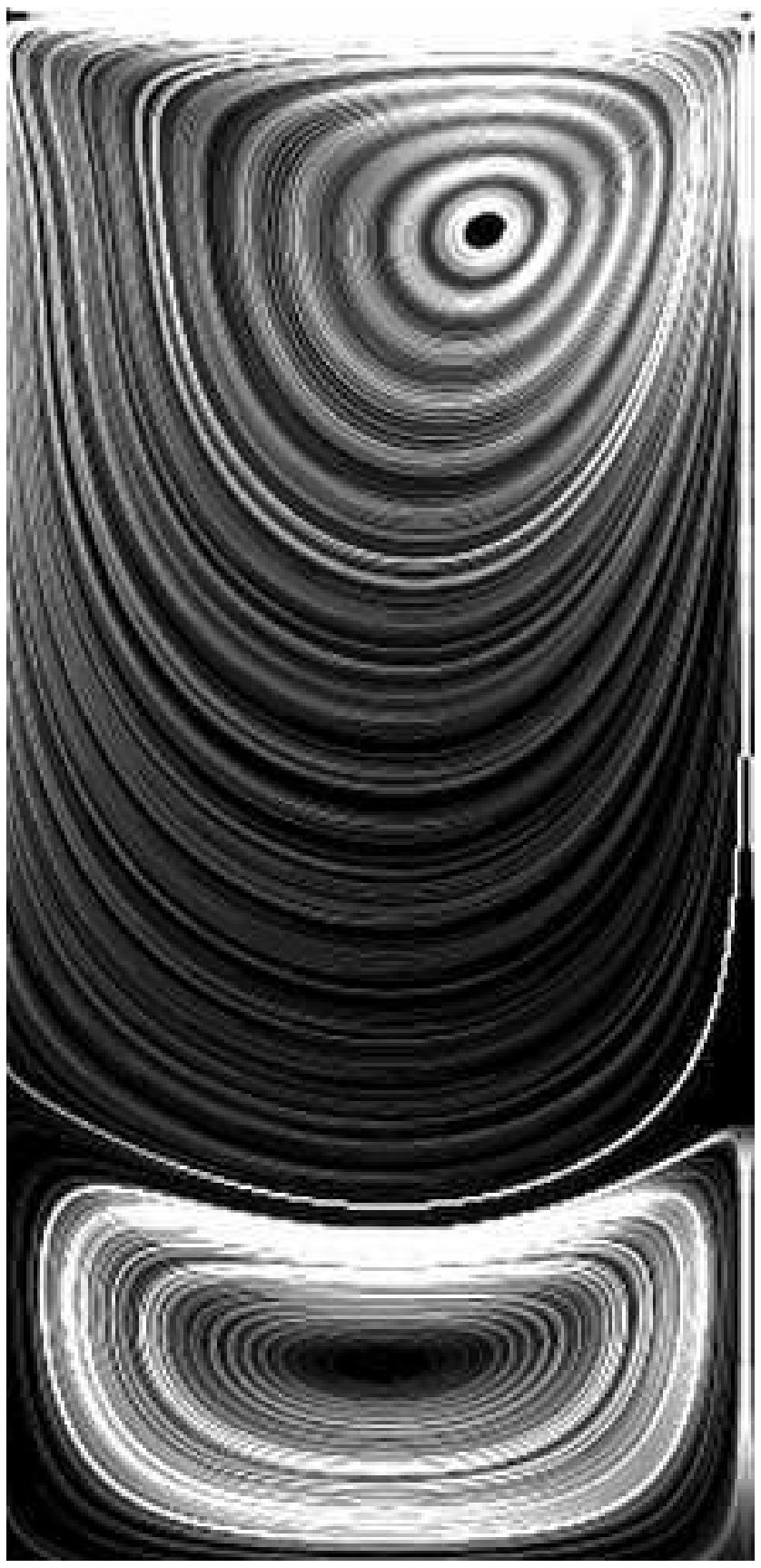

30x60, Aspect Ratio 1:2

Rysunek 10: Streamlines plot for SIMPLE (and SIMPLER - because they are the same) calculation of driven cavity with $Re = 100$ and different grid sizes and aspect ratios.

### 6.4 Convergence for SIMPLE and Vorticity-Sream algorthms

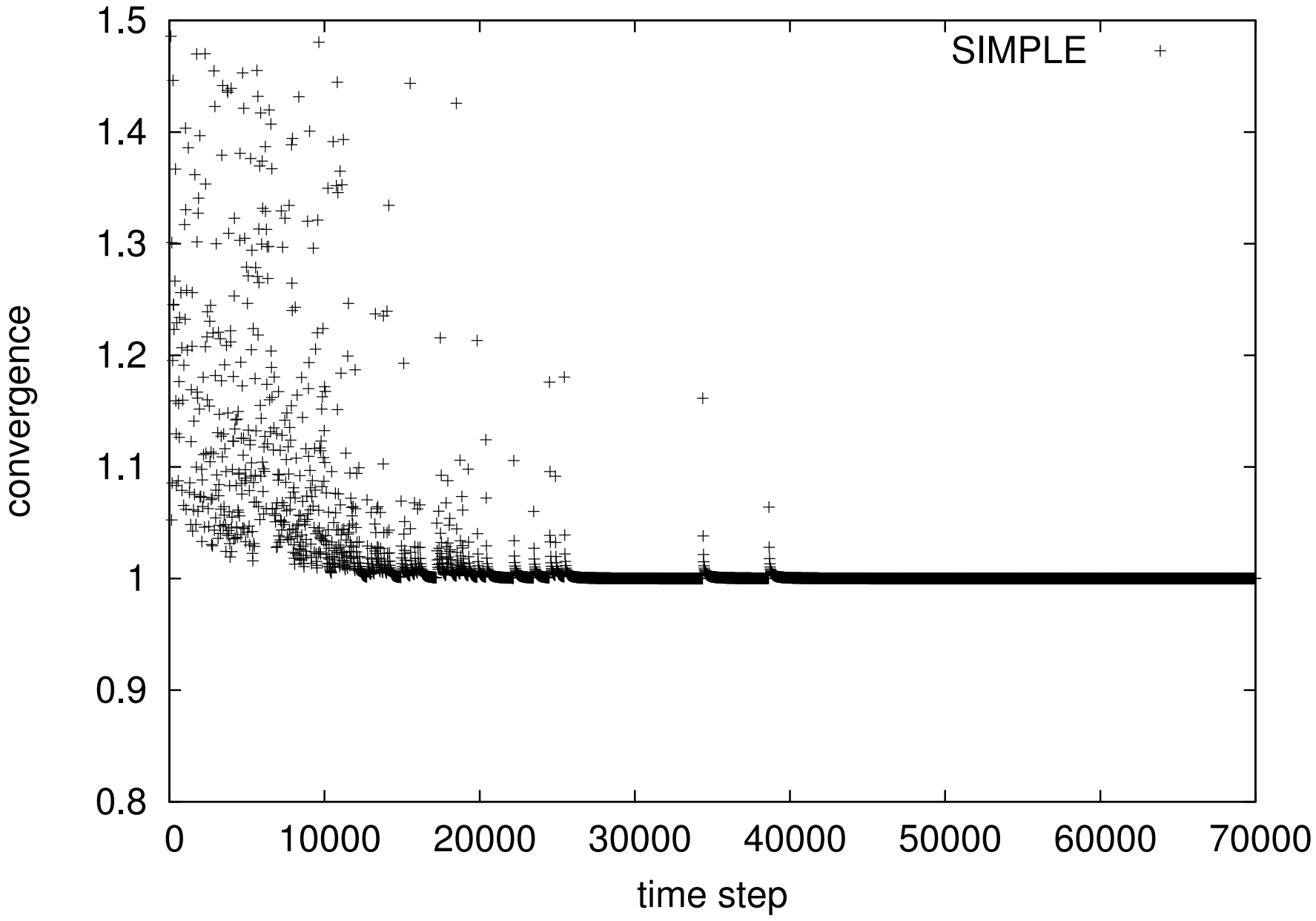

Rysunek 11: That figure shows how convergence changes during iteration steps. On $y$ axis we have $\frac{|v^{n+1}|}{|v^n|}$ variable.

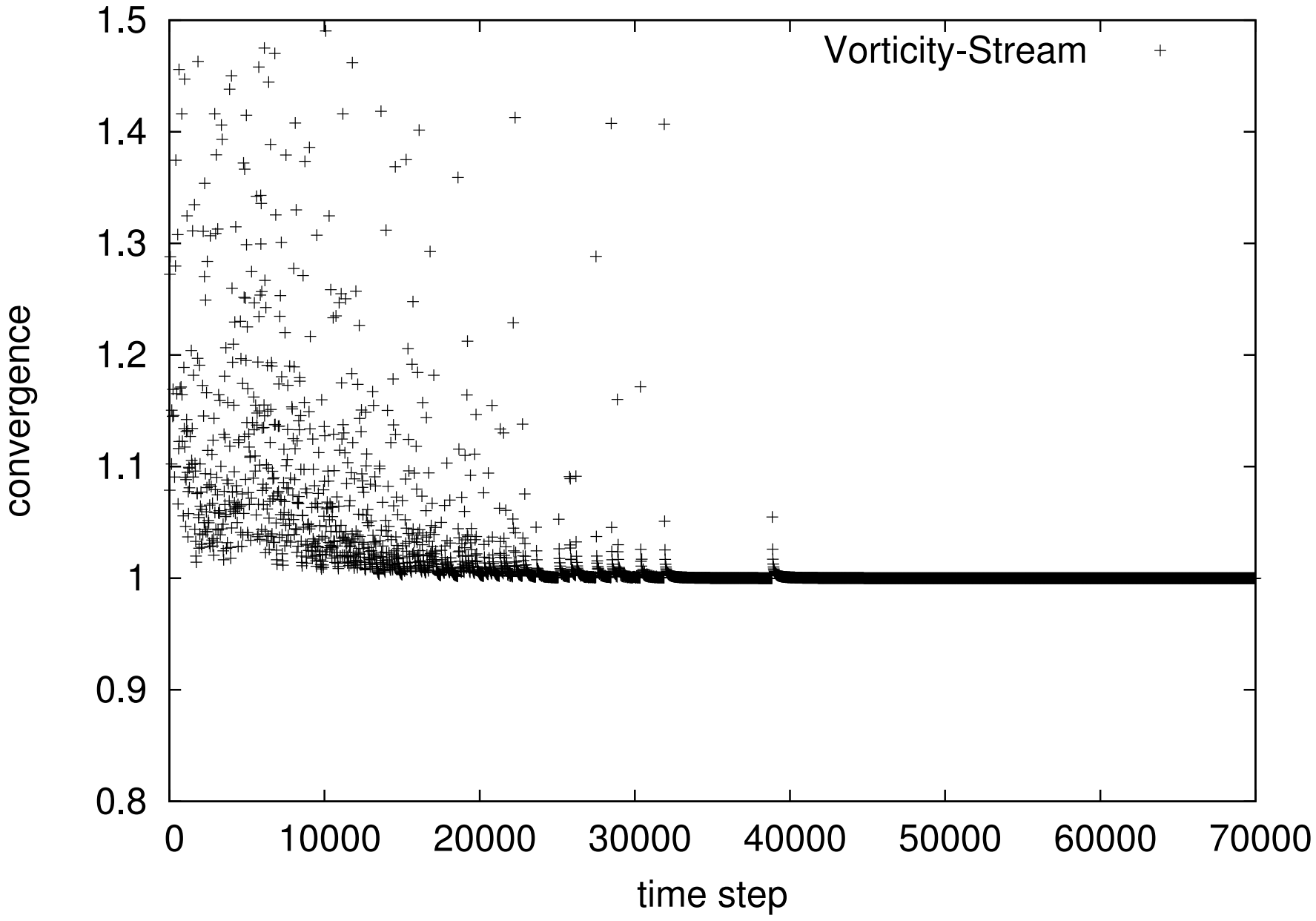

Rysunek 12: That figure shows how convergence changes during iteration steps. On $y$ axis we have $\frac{|v^{n+1}|}{|v^n|}$ variable.

### 6.5 Convergence comparision

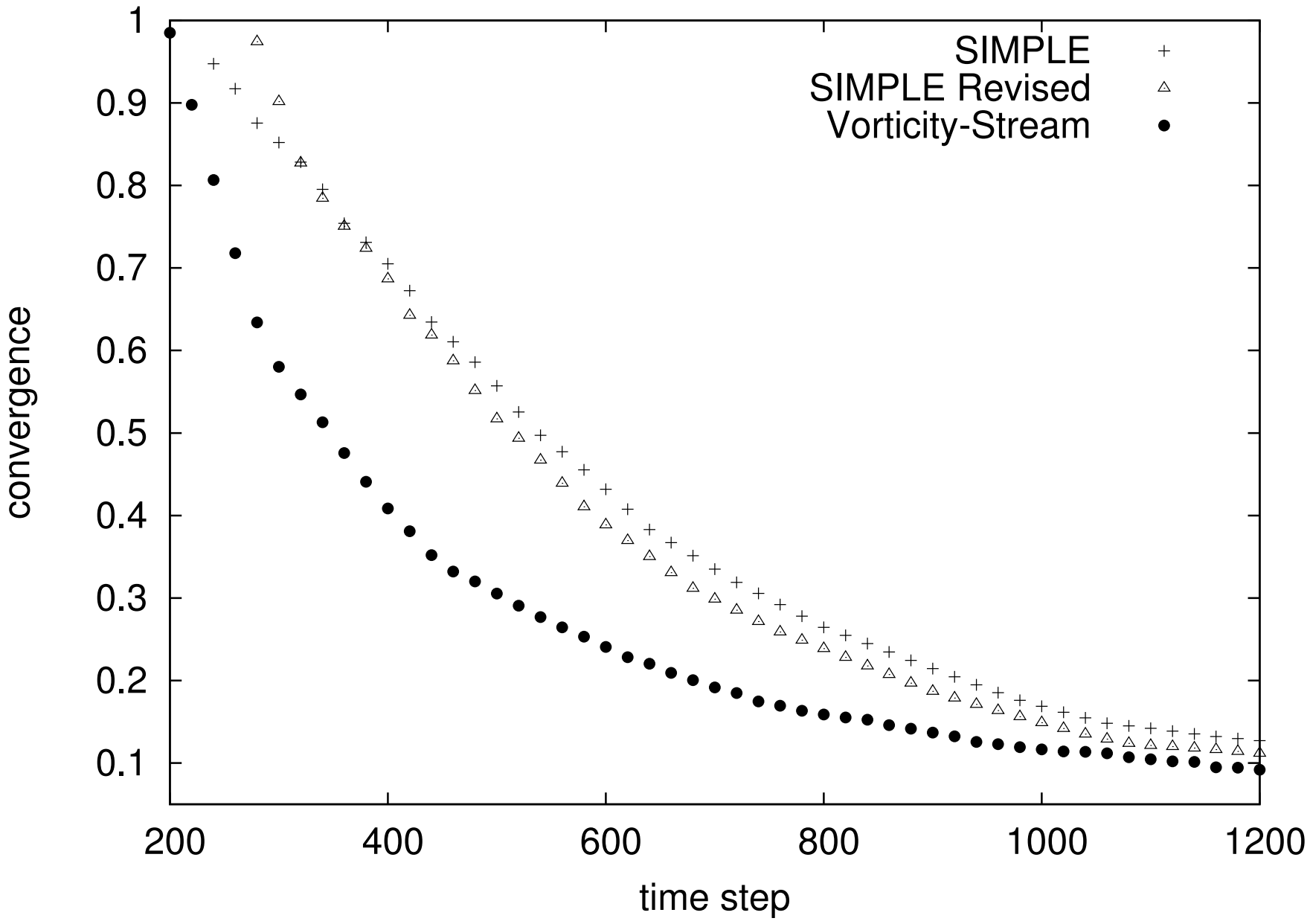

Rysunek 13: Convergence test for three solution algorithms. A lot of problems appeared there. Convergence speed depends on a lot of things, so for different properties of calculation (Reynolds numbers, spatial grid resolution, poisson equation accuracy etc.) different results appears. That results computed for $Re = 300$ and grid $30x30$ shows that Vorticity-Stream function solver converge faster than SIMPLER and SIMPLE. Anyway - more carefully study should be made there to make sure about that results. On the $y$ axis we have $|v^{n-1} - v^n|$ convergence coefficient.

# 7 Calculation For Flows over Obstacles

In that section I present some calculations made to test my SIMPLE solver for problems other than Driven Cavities. There were some problems with boundary conditions and still more work is needed there, but fortunately results are really nice.

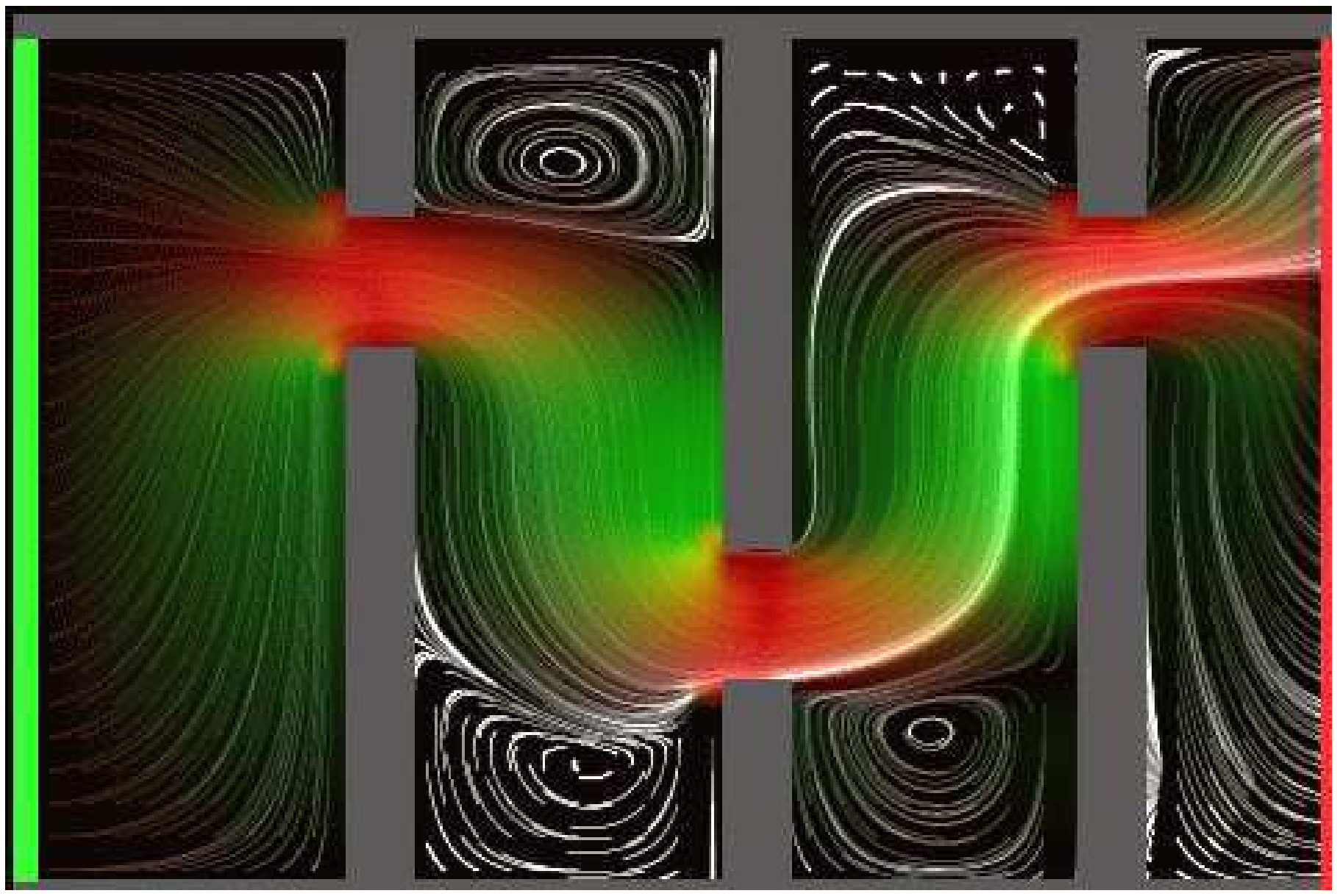

Rysunek 14: Flow of Incompressible fluid over set of holes. Calculation made for $Re = 250$ and grid size $60x40$.

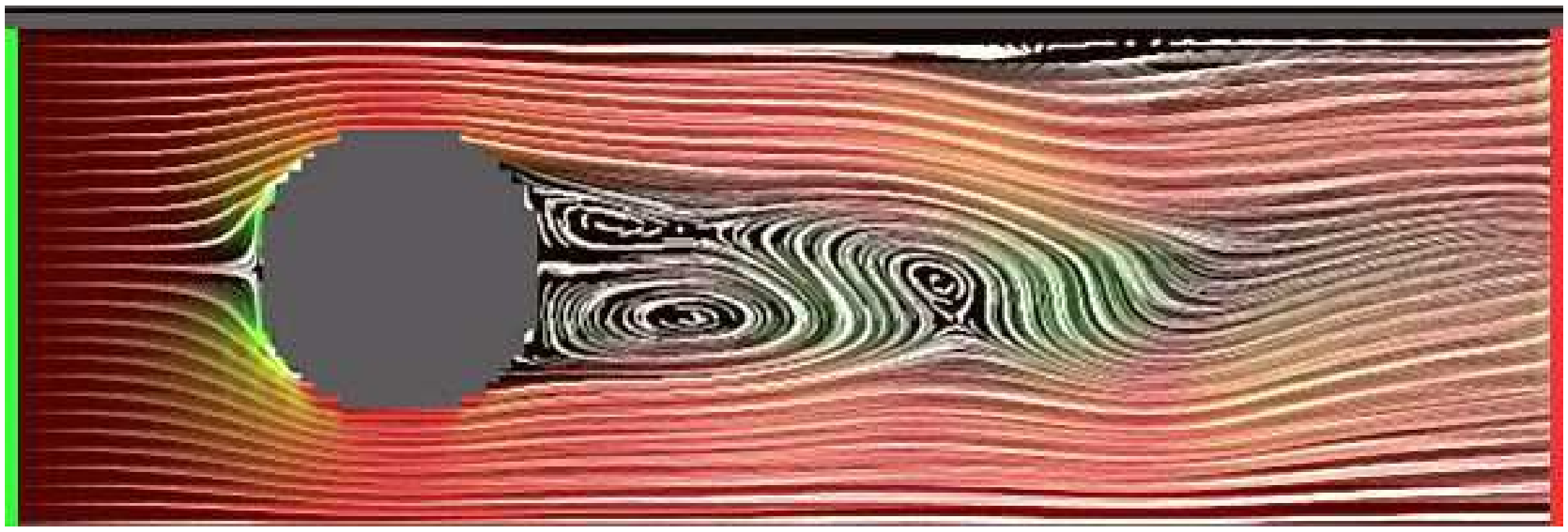

Rysunek 15: A Vortex-Karmann Street. Calculation made for $Re = 400$ and grid size $119x40$.

More results and an application "Hydrodynamica" for Windows operating system you can download free of a home page of an author.

# 8 Conclusion

I have developed three different methods for calculation of incompressible fluid flow. Tests for simple Driven Cavity problem were made. I compared convergence speed for all the methods and it seems that convergence speed depends on problem formulation and physical properties of simulated system. For future I will try to concern more on how to treat boundary conditions for both - pressure based and vorticity-stream function methods. Also some kind of user-friendly software will be released in near future. I would like to thank Grzegorz Juraszek for English language checking.

# A Appendix A

To calculate primed velocity correction values we use approximate forms of the momentum equations:

$$\frac{\partial u'}{\partial t} = -\frac{\partial p'}{\partial x} \tag{39}$$

$$\frac{\partial v'}{\partial t} = -\frac{\partial p'}{\partial y} \tag{40}$$

If we assume that velocity correction are zero at the previous time step, we can get straightforward expressions for velocity corrections at current time step:

$$u' = -\frac{1}{\Delta t}\frac{\partial p'}{\partial x} \tag{41}$$

$$v' = -\frac{1}{\Delta t}\frac{\partial p'}{\partial y} \tag{42}$$

Then, those two equations are discretized and we obtain simple expressions for calculation of velocity corrections:

$$u' = -\frac{1}{\Delta t \cdot \Delta x}(P'_{i+1,j} - P'_{i,j}) \tag{43}$$

$$v' = -\frac{1}{\Delta t \cdot \Delta y}(P'_{i,j+1} - P'_{i,j}) \tag{44}$$

# Literatura